\documentclass[
aps,
prd,
10pt,
notitlepage,
superscriptaddress,
nofootinbib,
numbers,
showpacs
]{revtex4-1}

\usepackage[utf8]{inputenc}
\usepackage[T1]{fontenc}
\usepackage{anyfontsize} 

\usepackage{amsmath,amssymb,amsfonts}
\usepackage{tensor,bm}  
\usepackage{mathrsfs} 

\usepackage{graphicx,subcaption}
\usepackage[dvipsnames]{xcolor}
\usepackage{orcidlink}

\usepackage{silence}
\usepackage{hyperref}
\hypersetup{
    colorlinks=true,
    citecolor=Purple,
    linkcolor=Purple,
    urlcolor=Purple,
    linktocpage=true,
    breaklinks=true
}

\usepackage[capitalize]{cleveref}

\begin{document}

\title{Dymnikova Black Hole Tidal Forces}

\author{M. H. Mac\^{e}do}
\email{matheus.macedo@fisica.ufc.br}
\affiliation{Universidade Federal do Cear\'a (UFC), Departamento de F\'isica,\\ Campus do Pici, Fortaleza - CE, C.P. 6030, 60455-760 - Brazil.}

%%%%%%%%%%%%%%%%%%%%%%%%%%%%%%%%%%%%%%%%%%%%%%%%%%%%%%%%%%%%%%%%%%%%%%

\author{A. A. M. Silva}
\email{anderson.alves@fisica.ufc.br}
\affiliation{Universidade Federal do Cear\'a (UFC), Departamento de F\'isica,\\ Campus do Pici, Fortaleza - CE, C.P. 6030, 60455-760 - Brazil.}

%%%%%%%%%%%%%%%%%%%%%%%%%%%%%%%%%%%%%%%%%%%%%%%%%%%%%%%%%%%%%%%%%%%%%%

\author{R. R. Landim}
\email{renan@fisica.ufc.br}
\affiliation{Universidade Federal do Cear\'a (UFC), Departamento de F\'isica,\\ Campus do Pici, Fortaleza - CE, C.P. 6030, 60455-760 - Brazil.}

\begin{abstract}
    In this work we investigate the tidal properties of the Dymnikova regular black hole and their effects on massive particles in radial free-fall. Starting from Dymnikova's static, spherically symmetric solution, we derive the equations governing timelike radial geodesics and construct an orthonormal tetrad adapted to a free-falling observer. We then obtain the radial and angular components of the tidal tensor and analyze their dependence on the black hole mass and the characteristic length scale of the de Sitter core. At large radial distances, tidal forces recover Schwarzschild behavior, whereas near the regular center, both components remain finite, reflecting the non-singular nature of the spacetime. We show that the radial and angular tidal forces vanish and change sign at characteristic radii inside the event horizon, indicating transitions between stretching and compression regimes. A particle released from rest outside the event horizon reaches a turnaround point located inside the Cauchy horizon, rather than reaching the regular center. We also solve the geodesic deviation equations for two sets of initial conditions and examine the evolution of the radial and transverse components of the deviation vector. Although the solutions asymptotically reproduce Schwarzschild behavior, they differ significantly in the inner region: the deviation vector components remain finite up to the turnaround point, whereas the corresponding radial component in the Schwarzschild case diverges at the singularity. These results demonstrate how the de Sitter core regularizes the tidal dynamics of extended falling bodies.
\end{abstract}


\maketitle
%%%%%%%%%%%%%%%%%%%%%%%%%%%%%%%%%%%%%%%%%%%%%%%%%%%%%%%%%%%%%%%%%%%%%%%%%%%%%%%
\section{Introduction}

Black holes occupy a central position in contemporary gravitational physics and astrophysics, serving as natural laboratories for testing gravity in the strong-field regime. The development of gravitational-wave astronomy and very-long-baseline interferometry has enabled these objects to be investigated through complementary observational channels. In particular, the Event Horizon Telescope (EHT) Collaboration reconstructed horizon-scale images of the supermassive compact objects M87*, located at the center of the giant elliptical galaxy M87 \cite{Akiyama_2019}, and Sagittarius A*, located at the center of the Milky Way \cite{EventHorizonTelescopeCollaboration_2022}. Complementarily, the first direct detection of gravitational waves from a binary black-hole merger by LIGO opened a new observational window for probing the dynamical strong-field regime of gravity \cite{PhysRevLett.116.061102}. These developments have considerably strengthened the motivation for investigating possible deviations from the classical black-hole geometries predicted by General Relativity.

The first exact static and spherically symmetric vacuum solution of Einstein's field equations was obtained by Schwarzschild in 1916 \cite{Schwarzschild:1916uq}. More general solutions were subsequently derived by including electric charge, leading to the Reissner--Nordström geometry \cite{Reissner,Nordstrom}, and angular momentum, leading to the Kerr solution \cite{Kerr:1963ud}. The combined effects of rotation and electric charge are described by the Kerr--Newman spacetime \cite{Newman1965}. These solutions form the classical family of asymptotically flat black holes in four-dimensional Einstein--Maxwell theory.

Despite their fundamental importance, these classical geometries contain curvature singularities in their maximal analytic extensions. In the Schwarzschild and Reissner--Nordström spacetimes, the curvature singularity occurs at ($r=0$), whereas in the Kerr and Kerr--Newman solutions it assumes the form of a ring located at ($r=0$) and ($\theta=\pi/2$). More generally, the singularity theorems establish that gravitational collapse leads, under broad physical and causal assumptions, to geodesic incompleteness \cite{Penrose1965,Hawking:1973uf}. Such behavior indicates a breakdown of the predictive domain of classical General Relativity and motivates the search for nonsingular gravitational configurations.

Early ideas toward singularity avoidance were proposed by Gliner \cite{Gliner1966} and Sakharov \cite{Sakharov1966}, who suggested that matter at extremely high densities could approach a vacuum-like state characterized by an effective negative pressure. These considerations motivated the replacement of the singular interior by a regular core. Within this context, Bardeen introduced the first widely recognized regular black-hole model \cite{Bardeen1968}. Its physical source was later interpreted by Ayón-Beato and García as a nonlinear magnetic monopole arising from General Relativity coupled to nonlinear electrodynamics \cite{AYONBEATO2000149}. Thus, the nonlinear-electrodynamics interpretation should be understood as a later completion of the original phenomenological model rather than as part of Bardeen's initial construction.

Several other regular black-hole geometries have since been proposed. Hayward developed a nonsingular model designed to describe the formation and evaporation of a black hole \cite{Hayward2006}, while Bambi and collaborators investigated rotating extensions of regular black-hole spacetimes \cite{BAMBI2013329}. Of particular interest is the solution introduced by Dymnikova \cite{Dymnikova1992}. This static and spherically symmetric geometry is supported by an anisotropic vacuum-like stress-energy tensor. It approaches the Schwarzschild solution at large radial distances, while its central region behaves as a de Sitter core. Consequently, the curvature singularity is replaced by a regular region characterized by a finite effective vacuum-energy density. Moreover, regular black hole remnants with such a de Sitter interior have been proposed as potential heavy dark matter candidates \cite{Dymnikova:2015yma}.

The Dymnikova geometry has recently been investigated from several complementary perspectives. Its possible emergence from an infinite tower of higher-curvature corrections has been analyzed in Ref.~\cite{KONOPLYA2024138945}, providing a theoretical framework for understanding the dynamical origin of the regular metric. Higher-dimensional extensions, together with their thermodynamic properties and quasinormal spectra, were studied in Ref.~\cite{Estrada:2024uuu,Paul:2023pqn,Macedo:2024dqb,Lutfuoglu:2025pzi,Errehymy:2026ftm}. The thermodynamics of a dynamical Dymnikova black hole, including a corrected formulation of the first law and quantum corrections, was considered in Ref.~\cite{Wu:2026uap,Ma:2024tqp,Ahmed:2026vce}. Generalizations involving a surrounding quintessence field \cite{Macedo:2025guc} and Einstein-Gauss-Bonnet gravity \cite{Errehymy:2025djk}  have also been proposed.

The phenomenological properties of Dymnikova-type geometries have been investigated through photon motion, unstable circular null orbits, black-hole shadows, and possible constraints from Event Horizon Telescope observations \cite{ERREHYMY2026140168}. Their dynamical response has also received considerable attention through the calculation of scalar and gravitational quasinormal modes \cite{Lutfuoglu:2025pzi,Macedo:2024dqb,Macedo:2025guc}. These studies demonstrate that the regularization scale may affect the oscillation frequencies, damping rates, optical structure, and thermodynamic behavior of the spacetime.

An additional and physically complementary way to probe a black-hole geometry is through the tidal forces experienced by freely falling observers. Tidal effects are determined locally by the spacetime curvature and are described operationally by the geodesic deviation equation. They provide information that is not directly contained in global observables such as shadows, quasinormal frequencies, or thermodynamic quantities. In regular black-hole spacetimes, the radial and angular tidal accelerations may remain finite, vanish at characteristic radii, or change from stretching to compression as an observer approaches the central region \cite{junior2020tidalforceschargedhayward}. General regularity criteria and selected properties of Jacobi fields and tidal forces have previously been discussed for nonsingular geometries, including the Dymnikova spacetime \cite{Maeda2022}. Nevertheless, a dedicated characterization of the radial and angular tidal sectors of the static Dymnikova black hole, emphasizing their critical points, sign changes, dependence on the regularization scale, and effects on an extended freely falling body, remains comparatively limited.

In this work, we investigate the tidal properties of the Dymnikova regular black hole. We first derive the equations governing radial timelike geodesics and construct an orthonormal frame adapted to a freely falling observer. We then obtain the radial and angular components of the tidal tensor and analyze their behavior in the exterior region, across the horizons, and toward the regular de Sitter core. Particular attention is devoted to the radii at which the tidal components vanish or change sign, as well as to their dependence on the parameters controlling the regularization of the geometry. Finally, we solve the corresponding geodesic deviation equations and examine the evolution of the radial and transverse components of the deviation vector. The results are compared with the Schwarzschild limit in order to determine how the replacement of the central singularity by a regular core modifies the stretching and compression experienced by an extended infalling body.

The paper is organized as follows. In Sec. II, we briefly present the Dymnikova solution and its main properties; in Sec. II A, we analyze radial geodesics. In Sec. III, we investigate the tidal forces, discussing their radial and angular components separately in Secs. III A and III B, respectively. In Sec. IV, we derive the equations governing the geodesic deviation vector and present the corresponding results for the radial and angular directions. Finally, in Sec. V, we summarize the main results and present our conclusions.

\section{Review of Dymnikova Black Hole}

\begin{figure}[h!]
    \centering
    \includegraphics[width=.62\linewidth]{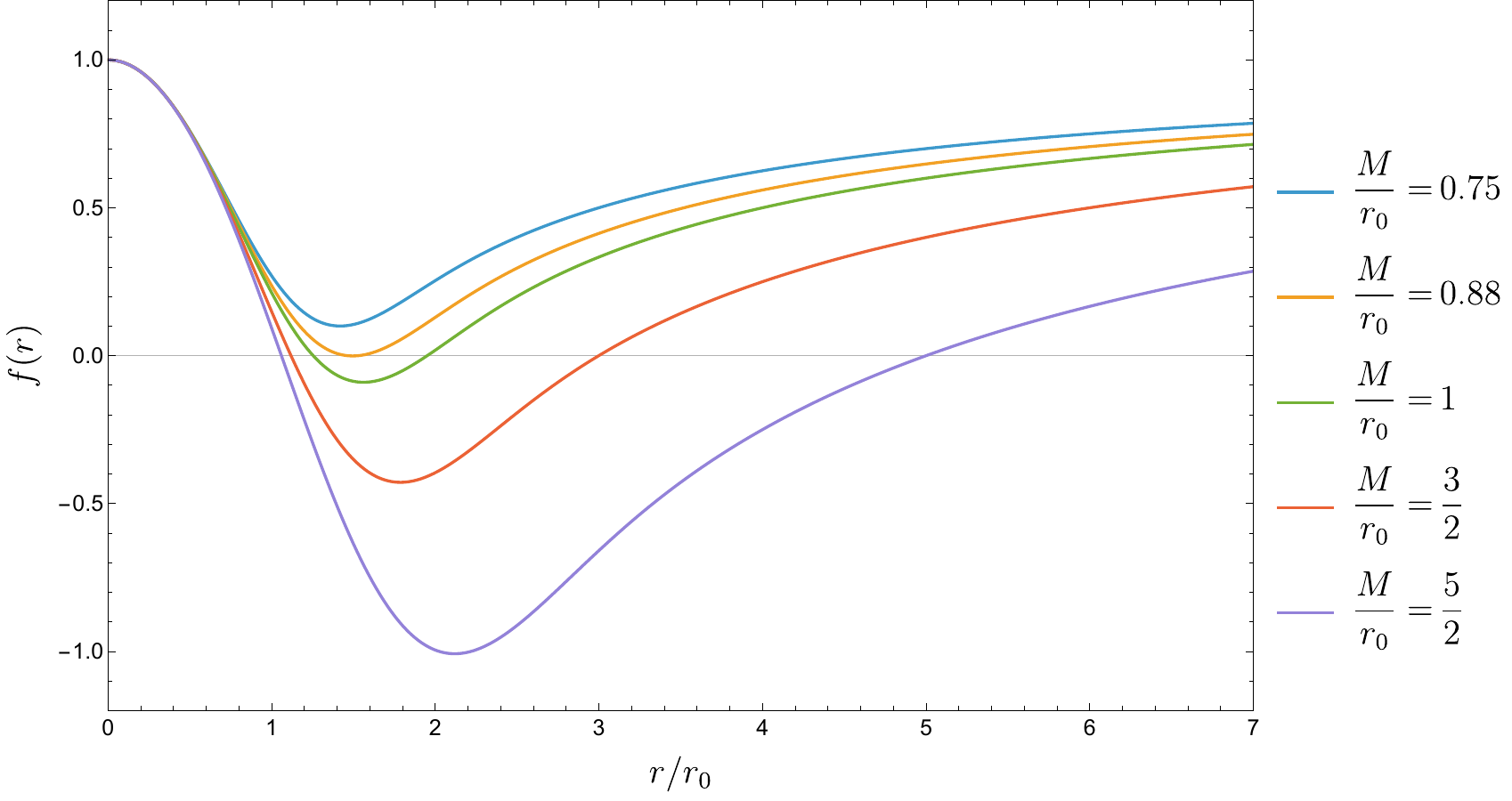}
    \caption{Plot of $f(r)$ as a function of $r/r_0$.}
    \label{Figure 1}
\end{figure}

In this section, we briefly discuss some of the features of the Dymnikova black hole. This spacetime represents a regular
solution of Einstein's field equations in which the Schwarzschild singularity is replaced by a de Sitter core. In Schwarzschild-like coordinates, the line element is given by \cite{Konoplya:2023aph}
\begin{equation}\label{2.1}
    ds^{2} = - f(r)\,dt^{2} + \frac{dr^{2}}{f(r)} + r^{2} d\Omega^{2},
\end{equation}
where
\begin{equation}
    d\Omega^{2} = d\theta^{2} + \sin^{2}\theta\,d\phi^{2},
\end{equation}
and the metric function is
\begin{equation}\label{2.3}
    f(r) =  1 - \frac{r_g}{r}\left(1 - e^{-r^3/r_{*}^3} \right).
\end{equation}
Here, $r_g = 2M$ denotes the Schwarzschild radius, where $M$ is the ADM mass and $r_{*}^3 = r_0^2r_g$ where $r_0$ determines the characteristic length scale of the de Sitter core \cite{Alencar:2023wyf}. For $r \gg r_{*}$, the metric reduces to the Schwarzschild black hole solution, whereas for $r \ll r_{*}$ we get the de Sitter black hole spacetime. The plot of equation (\ref{2.3}) is depicted in Fig.\ref{Figure 1} for various values of $r_g$. As we can see, the number of horizons is strongly dependent on the parameter values and that the two horizons, the inner and event, are quite close to each other.

The radial coordinates of the horizons are given by the zeros of the metric function $f(r)$, which leads to the transcendental equation $1 - \frac{2M}{r_h}\left[1 - \operatorname{exp}\left(-\frac{r_h^3}{2M r_0^2} \right) \right] = 0$. This equation cannot be solved analytically, but for $r_g \gg r_0$ we have two distinct horizons located at \cite{Macedo:2024dqb}
\begin{equation}
    r_{-} = r_0\left[1 - O\left(\operatorname{exp}\left(-\dfrac{r_0}{r_g} \right) \right) \right], \qquad \qquad r_{+} = r_g\left[1 - O\left(\operatorname{exp}\left(-\dfrac{r_g^2}{r_0^2} \right) \right) \right].
\end{equation}

The black hole mass, as a function of the horizon radius, is  equals to \cite{Macedo:2025guc}
\begin{equation}
    M(r_h) = \dfrac{r_h}{2\left[1 + \frac{r_0^2}{r_h^2}W\left(-\frac{r_h^2}{r_0^2}e^{-r_h^2/r_0^2} \right) \right]},
\end{equation}
where $W$ is the Lambert Function \cite{Lehtonen:2016}. Since $M(r_h)$ decreases for $r_0 < r_h < r_{crit}$ and increases for $r > r_{crit}$ it possesses a global minimum at $r_{crit} = 1.5\hspace{0.05cm} r_0$ and the corresponding critical mass is $M_{crit} = M(r_{crit}) = 0.88\hspace{0.05cm}r_0$. Therefore, horizons exist only for $M \geq M_{crit}$. For $M > M_{crit}$ the horizon equation admits two positive roots, $r_{-} < r_{crit} < r_{+}$ corresponding to the inner Cauchy horizon and outer event horizon, respectively. If $M = M_{crit}$ the two horizons coincide and for $M < M_{crit}$ no horizons are present.

\subsection{Radial Geodesics}

The radial motion of massive particles in the Dymnikova black hole spacetime follows directly from Eq. (\ref{2.1}), yielding
\begin{equation}\label{2.6}
    f(r)\dot{t}^2 - f^{-1}(r)\dot{r}^2 = 1,
\end{equation}
where the dot represents the differentiation with respect to the particle’s proper time $\tau$. Assuming purely radial motion, we impose $\dot{\theta} = \dot{\phi} = 0$. Furthermore, the timelike Killing vector implies the conserved energy per unit mass, $E = f(r)\dot{t}$, inserting this equation into Eq.(\ref{2.6}) we find that \cite{Idrissov:2025ugs}
\begin{equation}\label{2.7}
    E^2 = \dot{r}^2 + f(r).
\end{equation}
Consider a test particle released from rest at the radial position $b$, the conserved energy per unit rest mass is 
\begin{equation}
    E = \sqrt{f(b)} = \sqrt{1 - \dfrac{r_g}{b}\left(1 - e^{-b^3/r_{*}^3} \right) }
\end{equation}
A test particle released from rest at $b > r_{+}$ does not necessarily reach the central, but instead reaches a minimum radius $R_{stop}$ known as turning point, where the particle momentarily comes to rest before reversing its motion. The turning point is obtained by solving
\begin{equation}
    E^2 - f(R_{stop}) = 0.
\end{equation}
The turning point is always found inside the inner horizon. The Cauchy horizon is generally expected to be unstable as a consequence of the mass inflation mechanism \cite{PhysRevLett.63.1663, PhysRevD.41.1796, Simpson:1973ua}.
\begin{figure}[h!]
    \centering
    \includegraphics[width=.92\linewidth]{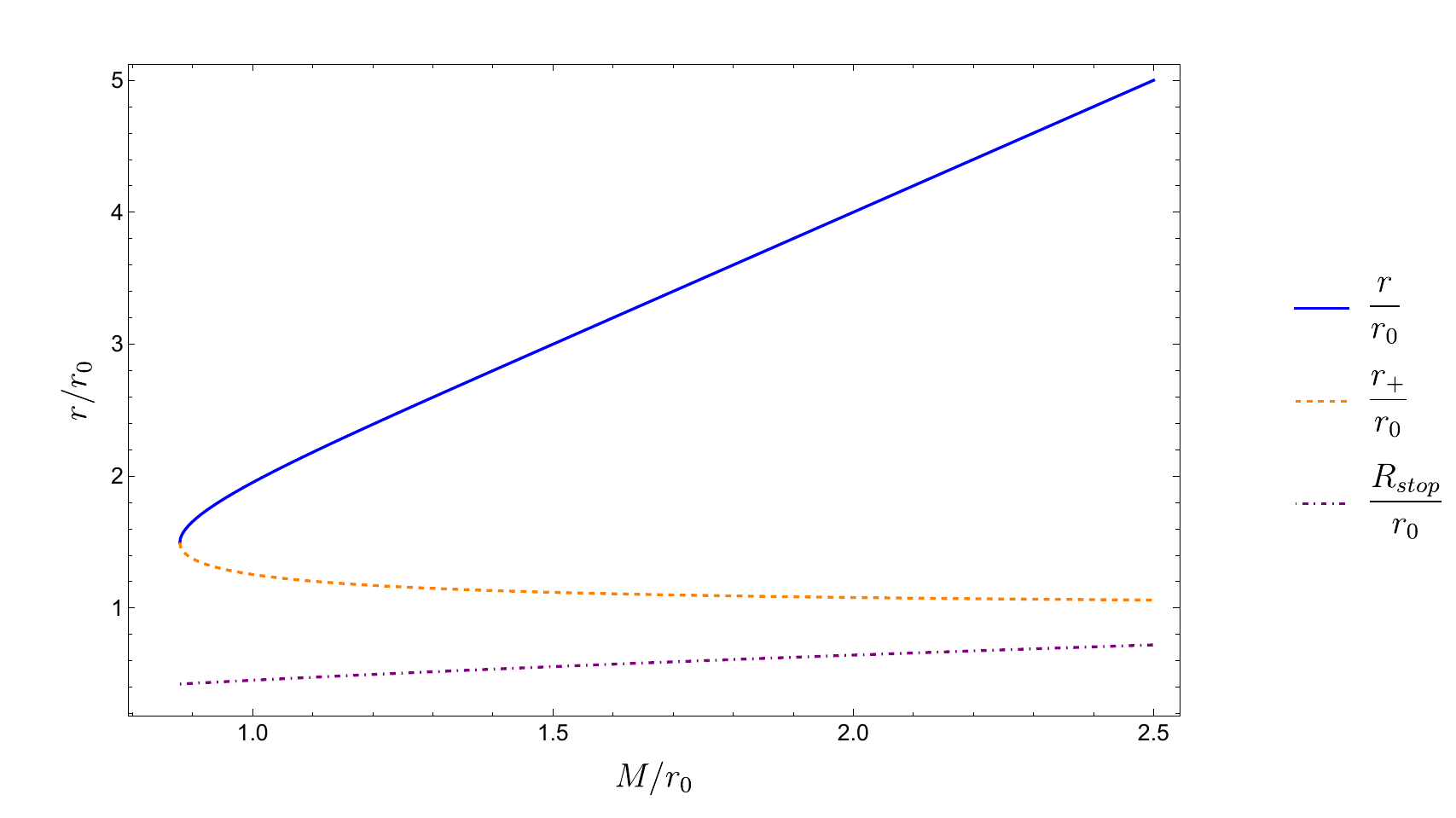}
    \caption{Plot of event horizon $r_{+}$, cauchy horizon $r_{-}$ and the turning point $R_{stop}$ as a function of the black hole mass $M$. We have chosen $b = 10r_0$ and $R_{stop}$ is always located inside the Cauchy horizon.}
    \label{Figure 2}
\end{figure}

\section{Tidal Forces}

 The relative acceleration between two infinitesimally separated particles is determined by the geodesic deviation equation \cite{Misner:1973prb}:
\begin{equation}\label{T.10}
    \dfrac{D^2\eta^{\hat{\alpha}}}{D\tau^2} = K\indices{^{\hat{\alpha}}_{\hat{\beta}}}\eta^{\hat{\beta}},
\end{equation}
where $\eta^{\hat{\alpha}}$ is the infinitesimal displacement vector between two nearby geodesics, and $K\indices{^{\hat{\alpha}}_{\hat{\beta}}}$ is the tidal tensor, given in terms of the Riemann tensor as 
\begin{equation}\label{T.11}
   K\indices{^{\hat{\alpha}}_{\hat{\beta}}} = R\indices{^{a}_{bcd}}e^{\hat{\alpha}}_{a}e^{b}_{\hat{0}}e^{c}_{\hat{0}}e^{d}_{\hat{\beta}} .
\end{equation}
The notation with hatted indices is used for the tetrad basis, while unhatted represent coordinate-basis components. We can project the geodesic deviation vector and the tidal tensor components in an orthonormal tetrad basis. We choose a tetrad basis attached to an observer in radial free fall in the Dymnikova black hole, which is given by:
\begin{equation}\label{T.12}
    \begin{aligned}
    & \hat{e}\indices{^{\mu}_{\hat{0}}} = \left(\dfrac{E}{f(r)}, -\sqrt{(E^2 - f(r)},0,0  \right), \\&
    \hat{e}\indices{^{\mu}_{\hat{1}} } = \left(-\dfrac{\sqrt{E^2 -f(r)}}{f(r)}, E,0,0  \right), \\&
    \hat{e}\indices{^{\mu}_{\hat{2}} } = \left(0, 0, \dfrac{1}{r},0  \right), \\&
    \hat{e}\indices{^{\mu}_{\hat{3}}} = \left(0, 0,0,\dfrac{1}{r\hspace{0.1cm}\sin{\theta}}  \right).
\end{aligned}
\end{equation}
These vector are mutually orthogonal and satisfy the Minkowski normalization condition
\begin{equation}
    \hat{e}\indices{^{\mu}_{\hat{\kappa}}}\hat{e}\indices{^{\nu}_{\hat{\sigma}}}g_{\mu\nu} = \eta_{\hat{\kappa}\hat{\sigma}},
\end{equation}
ensuring that a local observer measuring distances and angles in this frame obtains results consistent with special relativity. Using Eqs.(\ref{T.11}) and (\ref{T.12}) together with the Riemann tensor components of the Dymnikova black hole, we obtain the nonzero components of the tidal tensor as follows:
\begin{equation}\label{T.14}
     K\indices{^{\hat{r}}_{\hat{r}}} = \dfrac{r_{*}^3}{r_0^2r^3}\left[1 -e^{-r^3/r_{*}^3}\left(1+\frac{9r^6}{2r_{*}^6}\right) \right],  
\end{equation}
\begin{equation}\label{T.15}
     K\indices{^{\hat{\alpha}}_{\hat{\alpha}}} = -\dfrac{r_{*}^3}{2r_0^2r^3}\left[1 -  e^{-r^3/r_{*}^3}\left(1 + \frac{3r^3}{r_{*}^3} \right)\right],  
\end{equation}
where $\hat{\alpha} = (\hat{\theta}, \hat{\varphi})$. Combining Eqs.(\ref{T.14}) and (\ref{T.15}) with Eq. (\ref{T.10}), the relative acceleration of two nearby particles can be written as
\begin{equation}\label{T.16}
    \dfrac{D^2\eta^{\hat{r}}}{D\tau^2} = \dfrac{r_{*}^3}{r_0^2r^3}\left[1 -e^{-r^3/r_{*}^3}\left(1+\frac{9r^6}{8r_{*}^6}\right) \right]\eta^{\hat{r}},
\end{equation} 
\begin{equation}\label{T.17}
    \dfrac{D^2\eta^{\hat{\alpha}}}{D\tau^2} = -\dfrac{r_{*}^3}{2r_0^2r^3}\left[1 -  e^{-r^3/r_{*}^3}\left(1 + \frac{3r^3}{r_{*}^3} \right)\right]  \eta^{\hat{\alpha}}.
\end{equation}
 Eqs.(\ref{T.16}) and (\ref{T.17}) explicitly show how the tidal forces associated with the Dymnikova black hole are influenced by the parameters $M$ and $r_0$. For $r \gg r_{*}$, these expressions simplify to
\begin{equation}
    \left.\dfrac{D^2\eta^{\hat{r}}}{D\tau^2}\right|_{r\gg r_{*}} = \dfrac{2M}{r^3}\eta^{\hat{r}},
\end{equation}
\begin{equation}
    \left.\dfrac{D^2\eta^{\hat{\alpha}}}{D\tau^2}\right|_{r\gg r_{*}} = -\dfrac{M}{r^3}\eta^{\hat{\alpha}},
\end{equation}
which are the tidal forces for Schwarzschild black hole. Moreover, for $r \ll r_{*}$, we have
\begin{equation}
    \left.\dfrac{D^2\eta^{\hat{r}}}{D\tau^2}\right|_{r\ll r_{*}} = \dfrac{1}{r_0^2}\eta^{\hat{r}},
\end{equation}
\begin{equation}
    \left.\dfrac{D^2\eta^{\hat{\alpha}}}{D\tau^2}\right|_{r\ll r_{*}} = \dfrac{1}{r_0^2}\eta^{\hat{\alpha}},
\end{equation} 
representing the tidal forces for De Sitter spacetime. Observe that the  angular tidal force change from negative values for $r \ll r_{*}$ to positive values for $r \gg r_{*}$. This allows us to conclude that it must vanish at some radial coordinate $r$. Therefore, unlike the Schwarzschild black hole, the Dymnikova black hole can exhibit vanishing tidal forces.

\subsection{Radial Tidal Force}

As can be seen from Eq. (\ref{T.14}), the tidal force along the radial direction does not diverge at $r = 0$, instead it is given by
\begin{equation}
    \lim_{r \to 0}\dfrac{D^2\eta^{\hat{r}}}{D\tau^2} = \dfrac{1}{r_0^2}\eta^{\hat{r}}.
\end{equation}
\begin{figure}[h!]
    \centering
    \includegraphics[width=.82\linewidth]{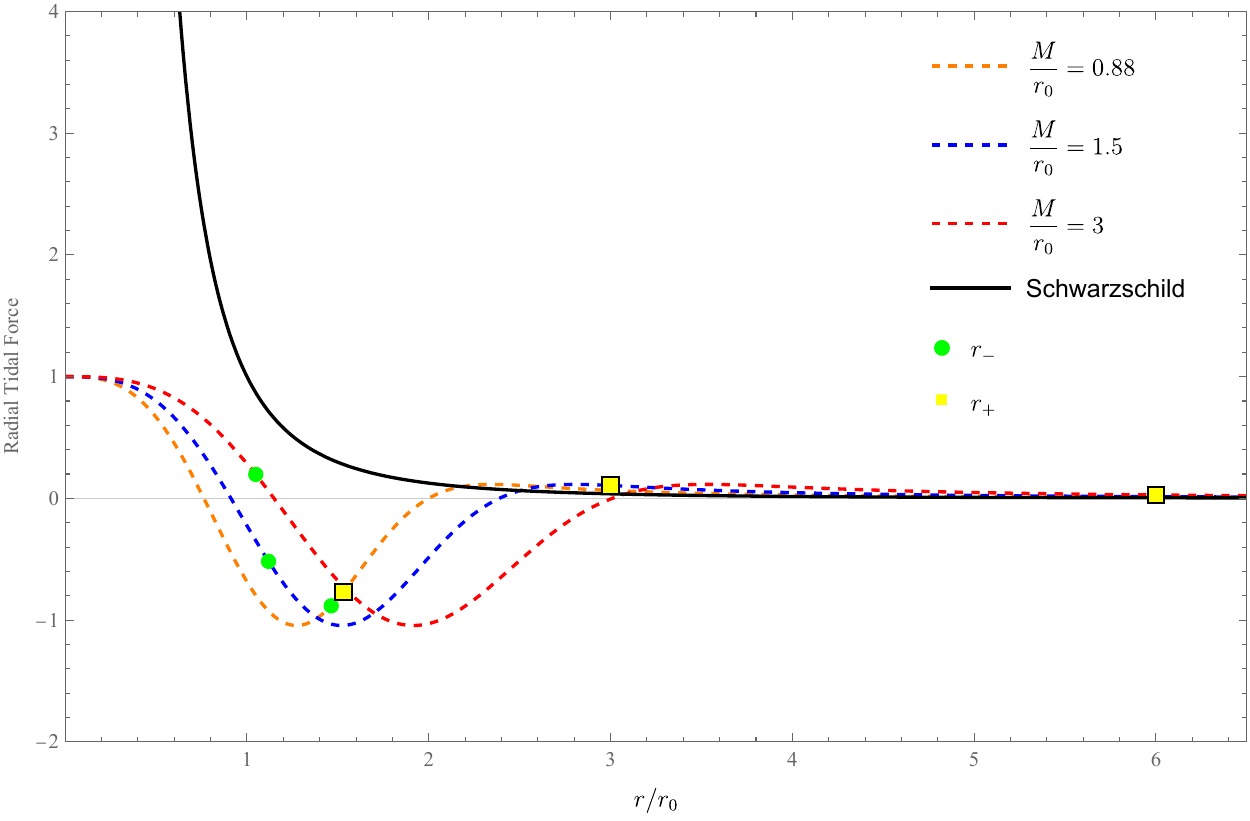}
    \caption{Radial tidal force for Dymnikova black holes with different choices of $M/r_0$, as well as for the Schwarzschild black hole. The positions of $R^{stop}, r_{-}$ and $r_{+}$ are exhibited in each plot.   }
    \label{Figure 3}
\end{figure}

In the Schwarzschild case the radial tidal force diverges at $r = 0$, the curvature singularity location. However, a radially infalling particle never reaches the origin, since it reverses its motion at the turning point $r = R_{stop}$. Furthermore, Eq.(\ref{T.16}) indicates that the radial tidal force may vanish at some radial coordinate. In Fig. \ref{Figure 3}, we plot the radial tidal force as a function of the radial coordinate. One can observe that the radial tidal forces vanish at a point located between the outer and inner horizons and for Schwarzschild black hole the radial tidal force remains positive and diverges, resulting in infinite radial stretching, as the body approaches the singularity.

\subsection{Angular Tidal Force}

Analogously to the radial tidal forces, the angular tidal forces not diverges at $r = 0$. From Eq.(\ref{T.17}), we find
\begin{equation}
     \lim_{r \to 0}\dfrac{D^2\eta^{\hat{\alpha}}}{D\tau^2} = \dfrac{1}{r_0^2}\eta^{\hat{\alpha}}.
\end{equation}
The behavior of the angular tidal forces as a function of the radial coordinate is shown in Figure 4 for different values of $M$ and revealing that the zero-tidal-force point, $r = R_{0}^{atf}$, lies outside the Cauchy horizon and inside the event horizon.
\begin{figure}[h!]
    \centering
    \includegraphics[width=.82\linewidth]{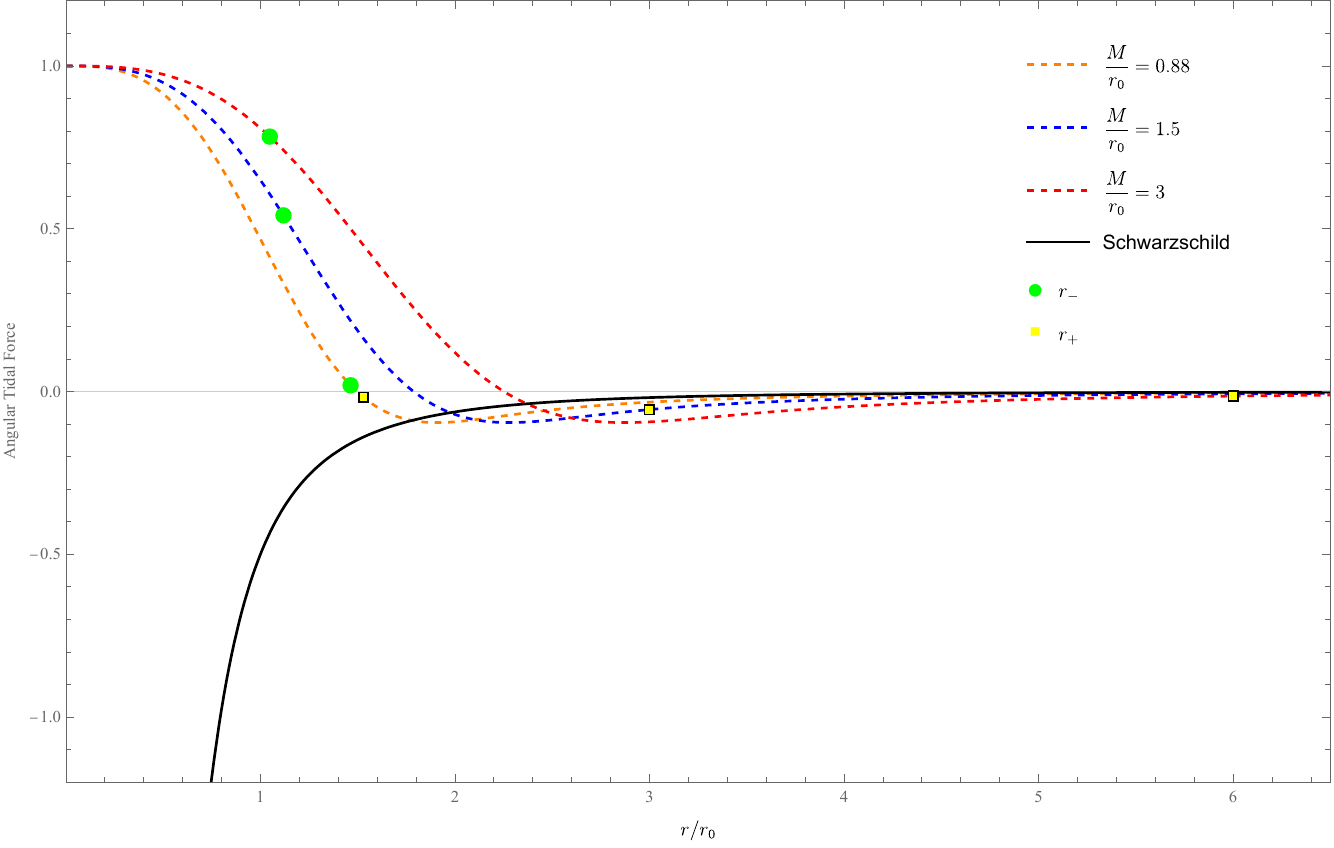}
    \caption{Angular tidal force for Dymnikova black holes with different choices of $M/r_0$, as well as for the Schwarzschild black hole. The positions of $R^{stop}, r_{-}$ and $r_{+}$ are exhibited in each plot. }
    \label{Figure 4}
\end{figure}

\section{Geodesic Deviation Vector}

The relation $dr/d\tau = -\sqrt{E^2 - f(r)}$, Eq. (\ref{2.7}), allows Eqs. (\ref{T.16})--(\ref{T.17}) to be rewritten as differential equations in $r$ and the geodesic deviation vectors to be obtained as functions of the radial coordinate. Therefore, for a test body formed by neutral dust particles accreting radially toward the Dymnikova black hole. From this, we obtain
\begin{equation}\label{T.24}
    (E^2 - f)\eta^{\hat{r}''} - \dfrac{f'}{2}\eta^{r'} + \dfrac{f''}{2}\eta^{\hat{r}} = 0,
\end{equation}
\begin{equation}\label{T.25}
    (E^2 - f)\eta^{\hat{\alpha}''} - \dfrac{f'}{2}\eta^{\hat{\alpha}'} + \dfrac{f''}{2r}\eta^{\hat{\alpha}} = 0.
\end{equation}
The geodesic deviation vector components, given by the solutions of Eqs.(\ref{T.24})-(\ref{T.25}), are
\begin{equation}
    \eta^{\hat{r}}(r) = \sqrt{E^2 - f}\left[C_1 + C_2\int\dfrac{dr}{(E^2 - f)^{3/2}} \right],
\end{equation}
\begin{equation}
    \eta^{\hat{\alpha}}(r) = \left[C_3 + C_4\int\dfrac{dr}{r^2(E^2 - f)^{1/2}} \right]r,
\end{equation}
where $C_1, C_2, C_3$ and $C_4$ are constants of integration. The integration constants are specified through two different sets of initial conditions:
\begin{equation}
    \eta^{\hat{\mu}}(b) > 0, \qquad \dot{\eta}^{\hat{\mu}}(b) = 0, \qquad \quad \text{(ICI),}
\end{equation}
\begin{equation}
    \eta^{\hat{\mu}}(b) = 0, \qquad \dot{\eta}^{\hat{\mu}}(b) > 0, \qquad \quad \text{(ICII),}
\end{equation}
for $b > r_{+}$. The first set of initial conditions (ICI) describes a collection of dust particles initially located at $r = b$, starting its motion from rest with all constituents sharing the same state of motion and no internal velocity dispersion. By contrast, the second set of initial conditions (ICII) represents a dust distribution that is ejected outward from $r = b$. In the following section we make a detailed study of the radial and transverse components of the geodesic deviation vector associated with each of these initial configurations.

\begin{figure*}[ht!]
\centering

%--------------------- Coluna da esquerda ---------------------%
\begin{minipage}[t]{0.48\textwidth}
\centering

\begin{subfigure}{\linewidth}
    \centering
    \includegraphics[width=\linewidth]{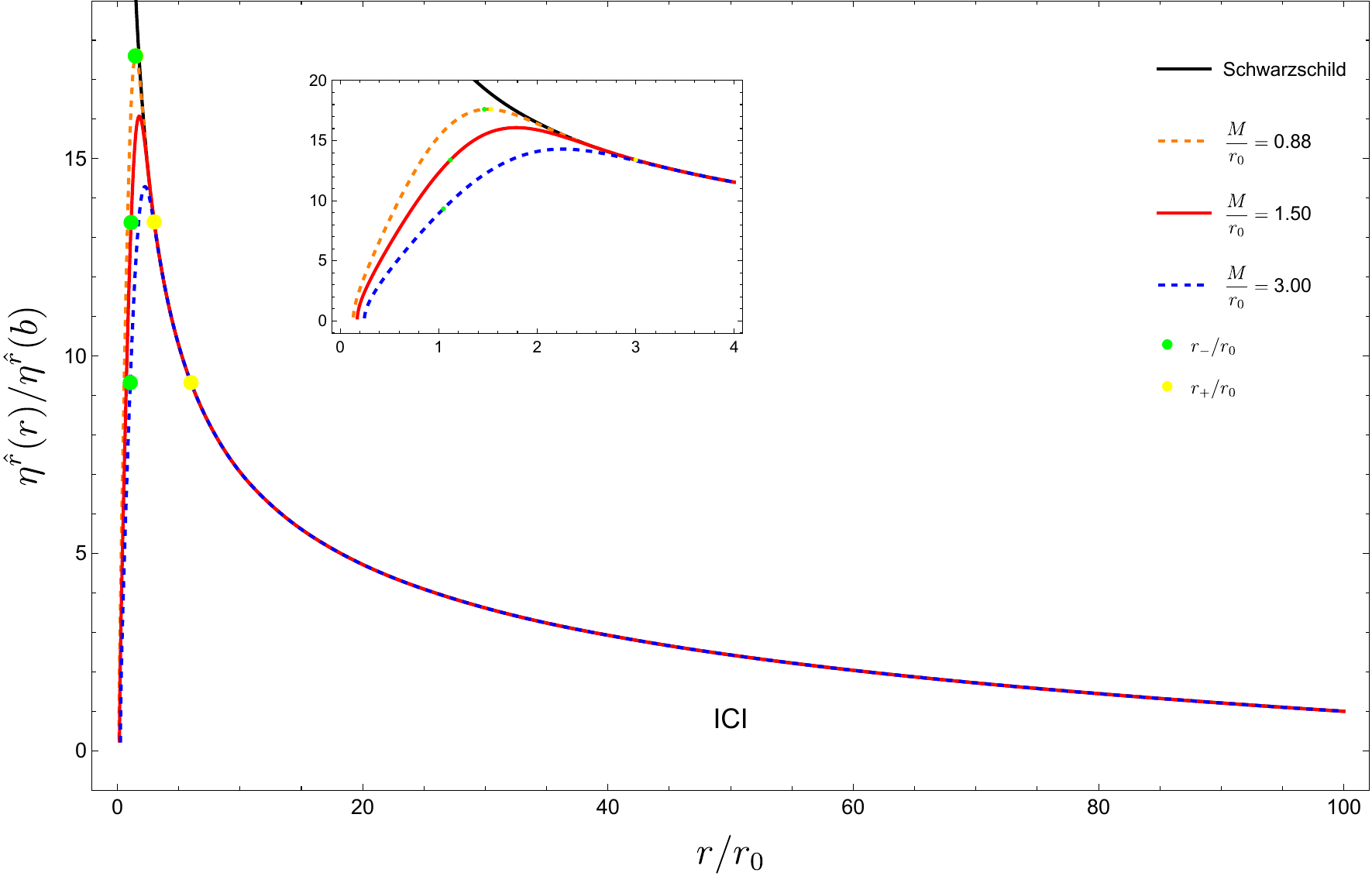}
\end{subfigure}

\vspace{0.3cm}

\begin{subfigure}{\linewidth}
    \centering
    \includegraphics[width=\linewidth]{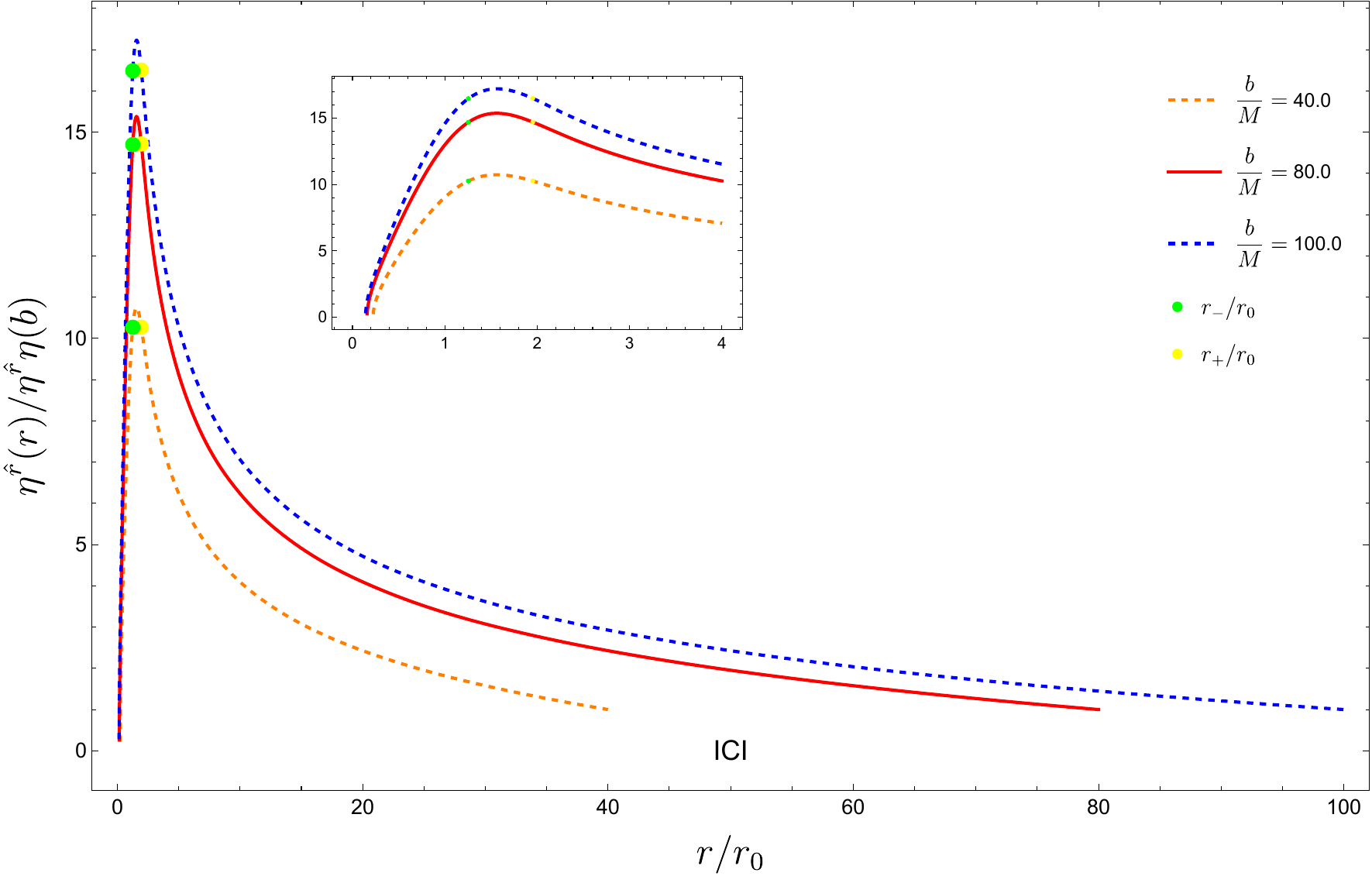}
\end{subfigure}

\caption{Radial component of the geodesic deviation vector of the Dymnikova black hole as a function of $r$. The upper panel shows the behavior for different values of $M/r_0$ with $b=100r_0$, while the lower panel corresponds to different values of $b$ with $M=r_0$, both for the initial condition ICI.}
\label{Figure 5}

\end{minipage}
\hfill
%--------------------- Coluna da direita ---------------------%
\begin{minipage}[t]{0.48\textwidth}
\centering

\begin{subfigure}{\linewidth}
    \centering
    \includegraphics[width=\linewidth]{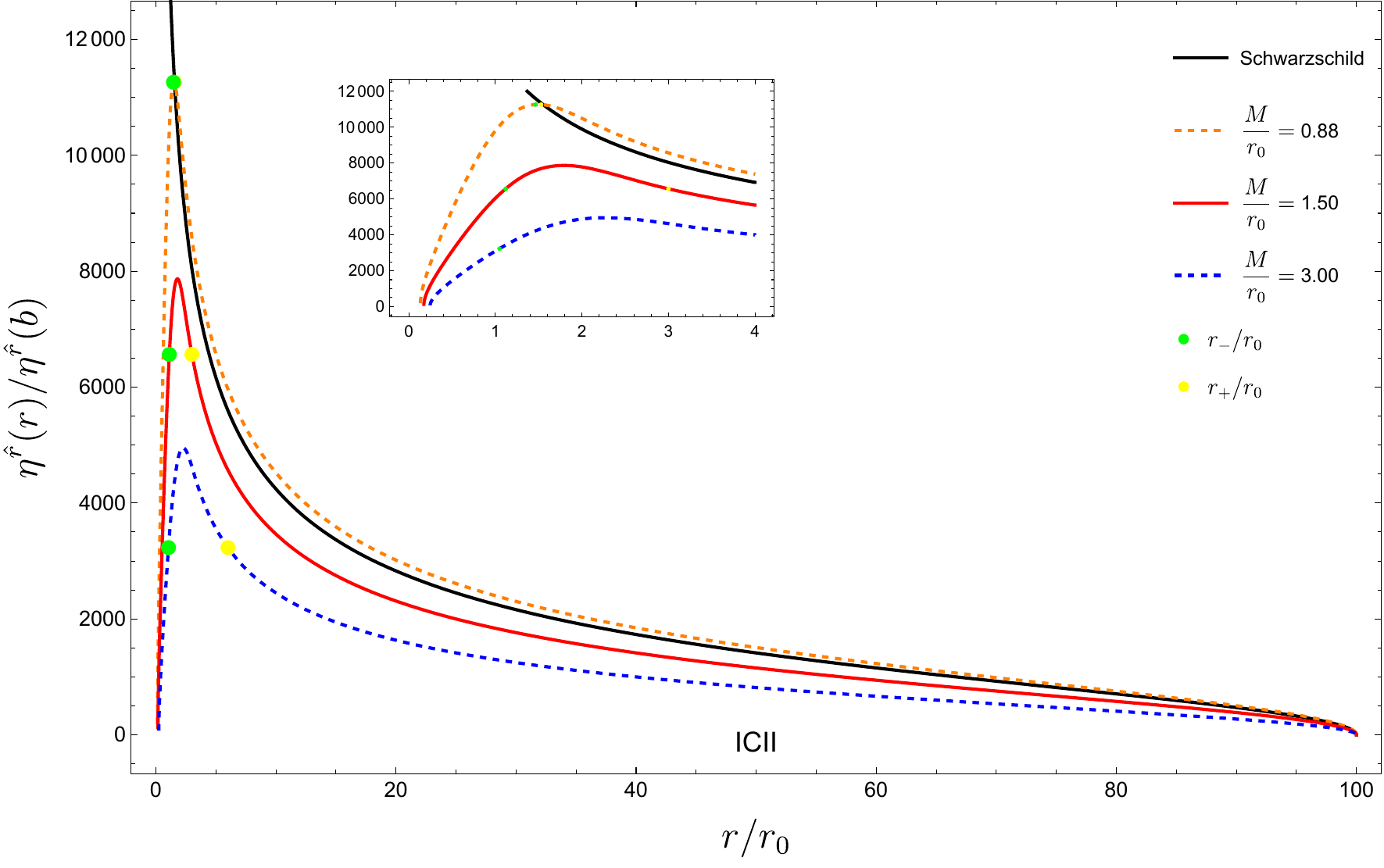}
\end{subfigure}

\vspace{0.3cm}

\begin{subfigure}{\linewidth}
    \centering
    \includegraphics[width=\linewidth]{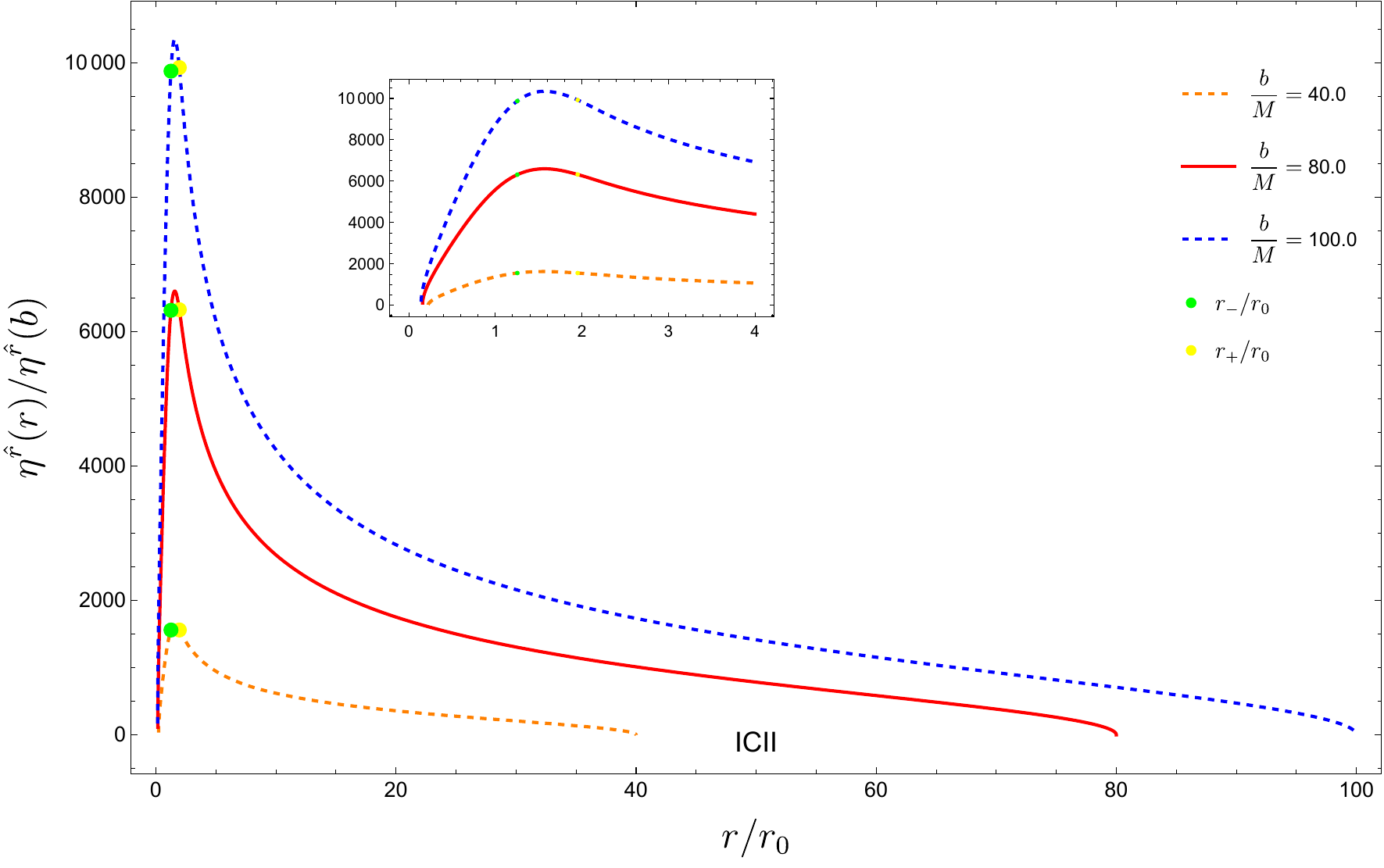}
\end{subfigure}

\caption{As in Fig.~\ref{Figure 5}, we plot the radial component of the geodesic deviation vector of the Dymnikova black hole as a function of $r$. The upper panel corresponds to different values of $M/r_0$ with $b=100r_0$, while the lower panel shows different values of $b$ with $M=r_0$, now for the initial condition ICII.}
\label{Figure 6}

\end{minipage}

\end{figure*}

Figure \ref{Figure 5} and \ref{Figure 6} shows the radial component of the geodesic deviation vector obtained from Eq. (\ref{T.24}) with the initial condition ICI and ICII. The upper panel displays the results for different values of $M/r_0$ with $b = 100r_0$, while the lower panel corresponds to different values of b with $M = r_0$. In both cases, the curves almost coincide at large distances from the black hole, showing that the asymptotic behavior is weakly affected by the parameters. As the particle falls radially inward from $r = b$, the deviation vector grows outside the event horizon, attains a maximum between the event and Cauchy horizons, and subsequently decreases until the stopping radius $R_{stop}$, a direct consequence of the change in sign of the radial tidal force. We also observe that, unlike the Dymnikova black hole, the radial component of the geodesic deviation vector diverges at the central singularity in the Schwarzschild spacetime.

\begin{figure*}[ht!]
\centering

%--------------------- Coluna da esquerda ---------------------%
\begin{minipage}[t]{0.48\textwidth}
\centering

\begin{subfigure}{\linewidth}
    \centering
    \includegraphics[width=\linewidth]{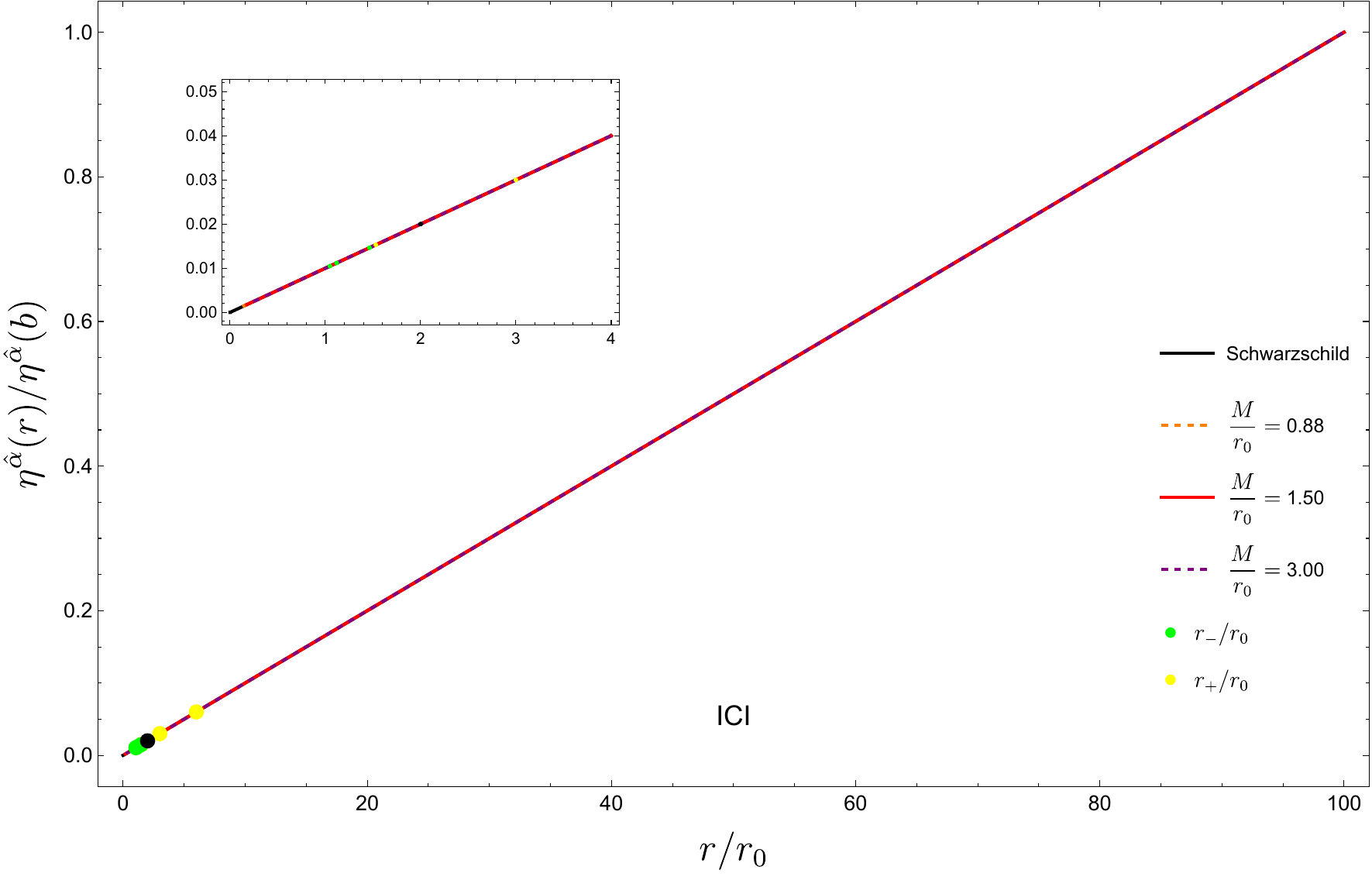}
\end{subfigure}

\vspace{0.3cm}

\begin{subfigure}{\linewidth}
    \centering
    \includegraphics[width=\linewidth]{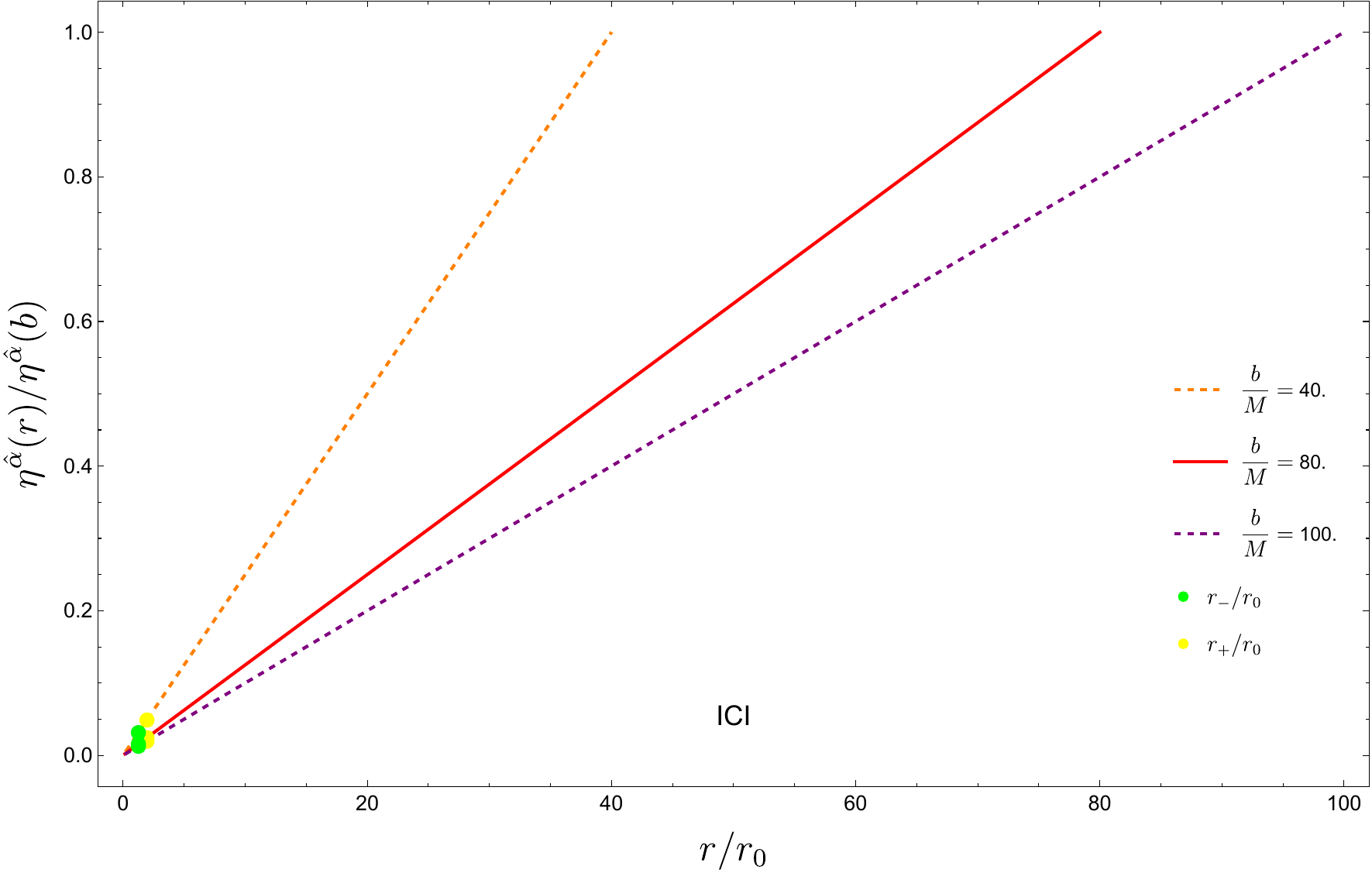}
\end{subfigure}

\caption{Angular component of the geodesic deviation vector of the Dymnikova black hole as a function of $r$. The upper panel shows the behavior for different values of $M/r_0$ with $b=100r_0$, while the lower panel corresponds to different values of $b$ with $M=r_0$, both for the initial condition ICI.}
\label{Figure 7}

\end{minipage}
\hfill
%--------------------- Coluna da direita ---------------------%
\begin{minipage}[t]{0.48\textwidth}
\centering

\begin{subfigure}{\linewidth}
    \centering
    \includegraphics[width=\linewidth]{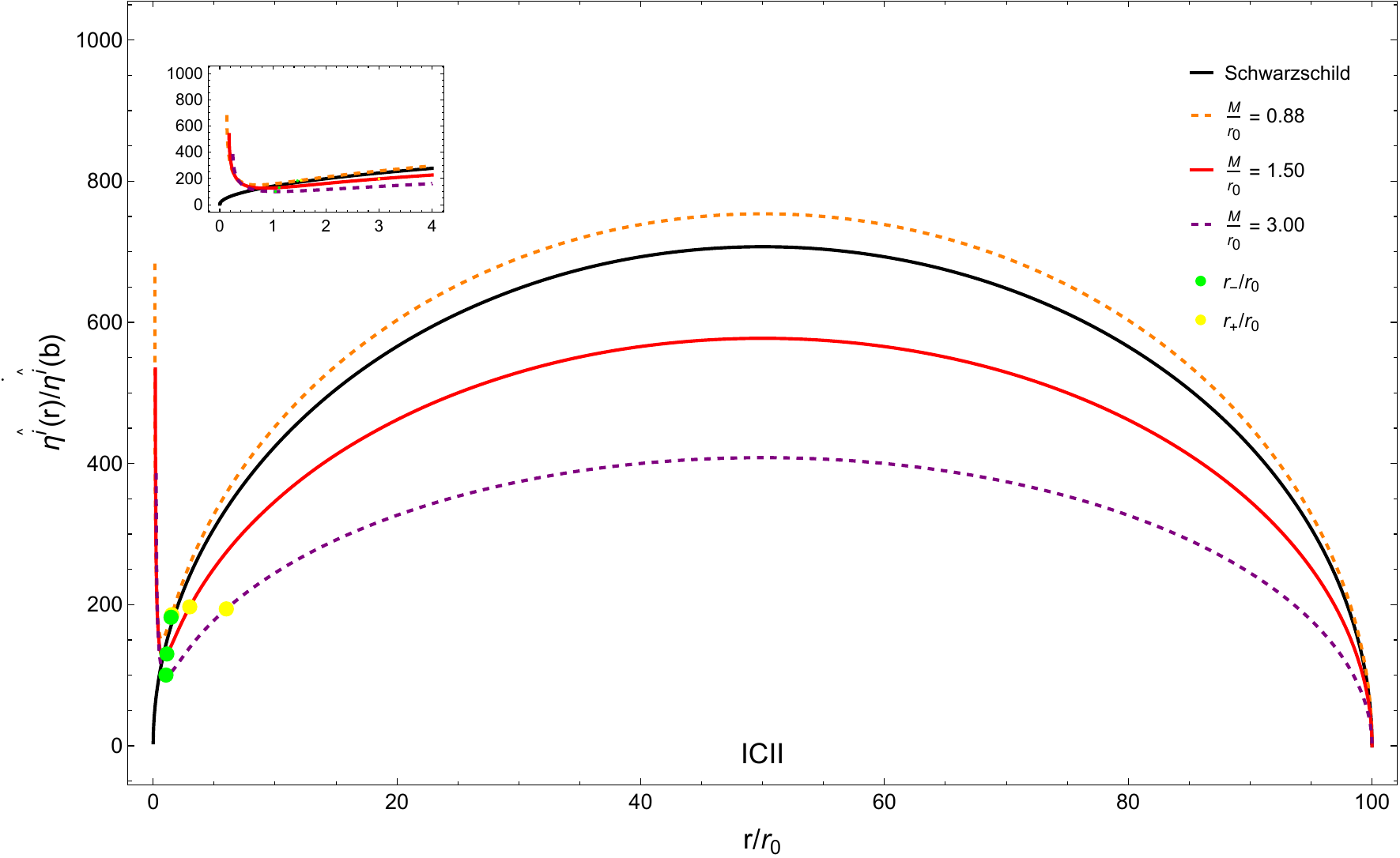}
\end{subfigure}

\vspace{0.3cm}

\begin{subfigure}{\linewidth}
    \centering
    \includegraphics[width=\linewidth]{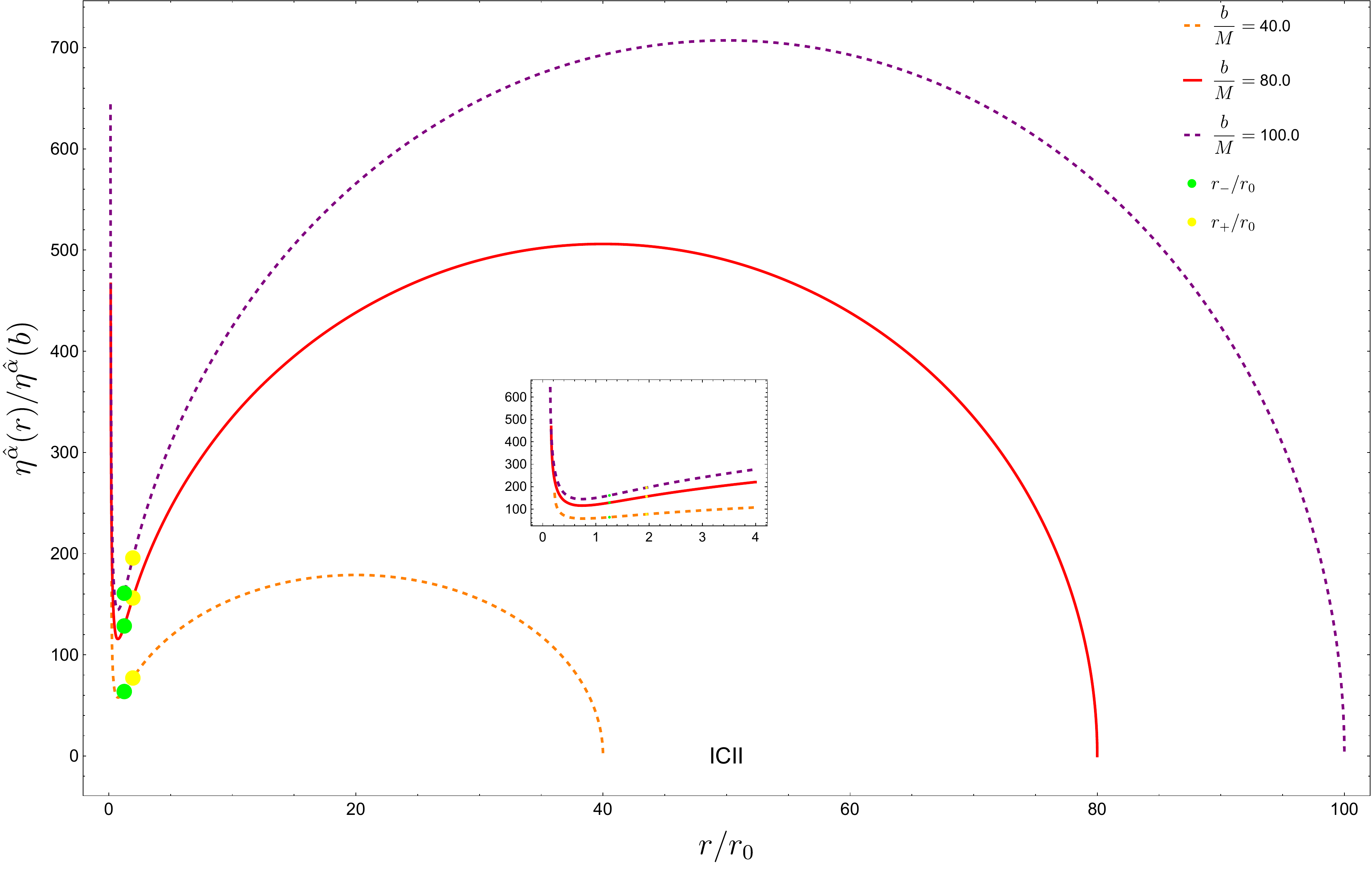}
\end{subfigure}

\caption{As in Fig.~\ref{Figure 7}, we plot the angular component of the geodesic deviation vector of the Dymnikova black hole as a function of $r$. The upper panel corresponds to different values of $M/r_0$ with $b=100r_0$, while the lower panel shows different values of $b$ with $M=r_0$, now for the initial condition ICII.}
\label{Figure 8}

\end{minipage}

\end{figure*}

In Fig. \ref{Figure 7}, we plot the angular component of the geodesic deviation vector obtained by solving Eq.(\ref{T.25}) with the initial condition ICI. The upper panel corresponds to different values of $M/r_0$ with $b = 100r_0$, while the lower panel shows different values of $b$ with $M = r_0$. We observe that, far from the black hole, the angular component of the geodesic deviation vector exhibits nearly the same behavior for all parameter values, reflecting the asymptotically Schwarzschild nature of the Dymnikova spacetime and decreases linearly with $r$, as expected, since the geodesic motion is purely radial.

In Fig. \ref{Figure 8}, we plot the angular component of the geodesic deviation vector obtained by solving Eq.(\ref{T.25}) with the initial condition ICII. As in the previous case, the upper panel corresponds to different values of $M/r_0$ with $b = 100r_0$, while the lower panel shows different values of $b$ with $M = r_0$. The angular component initially increases, reaches a maximum approximately halfway along the trajectory, and then decreases as the particle approaches the black hole. After reaching a minimum inside the event horizon, it increases again before the particle reaches the stopping radius, $R_{stop}$. This behavior reflects the change in sign of the angular tidal force.

\section{Conclusion}

In this paper, we have examined the tidal forces in the spacetime of the Dymnikova regular black hole. Our results demonstrate that both the radial and angular tidal forces remain bounded across the entire spacetime, including at the regular center. We found that, depending on the value of the mass parameter $M/r_0$, the radial tidal force reverses sign within the event horizon, indicating a transition from stretching to compressive behavior. Similarly, the angular tidal force also changes sign, leading to a vanishing tidal-force radius located between the Cauchy and event horizons.

We further investigated the geodesic deviation equations for radially infalling under two different sets of initial conditions.  In all cases, the geodesic deviation vector exhibits a qualitative behavior similar to that found in the Schwarzschild spacetime far from the black hole, reflecting asymptotically Schwarzschild character of the Dymnikova solution. Nevertheless, the behavior changes significantly within the event horizon. While the radial component of the geodesic deviation vector diverges at the Schwarzschild singularity, all components remain finite throughout the Dymnikova spacetime as a direct consequence of its regular core.

The radial geodesic deviation vector was found to increase during the first stage of the infall, reaching its peak inside the event horizon. Afterward, it decreases toward the stopping radius, $R_{stop}$, as a consequence of the sign reversal of the radial tidal force. The angular component displays a complementary behavior, it either decreases continuously or first increases to a maximum before declining, in agreement with the sign change of the angular tidal force.   In every case, the geodesic deviation vector remains finite up to the stopping radius, highlighting the absence of a physical singularity in the Dymnikova spacetime.

\acknowledgments  
\hspace{0.5cm} The authors acknowledge Coordena\c{c}\~{a}o de Aperfei\c{c}oamento de Pessoal de N\'{i}vel Superior (CAPES) and the Conselho Nacional de Desenvolvimento Cient\'{i}fico e Tecnol\'ogico (CNPq).
%==========================================

%==========================================
\bibliographystyle{apsrev4-1}
\bibliography{ref}
%==========================================
\end{document}